# Non-Hermitian reconstruction of hybridized hyperbolic shear polaritons

Zijia Dang[1], Zhe Cui[1], Gang Zhong[1], Jingying Liu[1,2], Qingdong Ou[1,2]*

[1]Macao Institute of Materials Science and Engineering (MIMSE), Sino-Luso Joint Laboratory for Optoelectronics, Macau University of Science and Technology, Taipa, Macao 999078, China.
[2]Macau University of Science and Technology Zhuhai MUST Science and Technology Research Institute, Zhuhai 519031, China

Correspondence:
Qingdong Ou (qdou@must.edu.mo)

**Abstract:**
Hyperbolic shear polaritons (HShPs) represent a distinctive class of anisotropic polaritonic states that enable directional subwavelength light manipulation through asymmetric momentum-space dispersion, while their control in twisted systems has so far relied mainly on geometric and spectral parameters. Here, we introduce non-Hermitian dissipation as an additional degree of freedom for tuning HShPs in twisted van der Waals heterostructures. In twisted α-$MoO_3$/α-$V_2O_5$ heterostructures with closely matched isofrequency contours but distinct dielectric losses, dissipation modifies the complex hybrid polaritonic states and thereby tunes the resulting momentum-space and real-space asymmetry. Near-field infrared imaging combined with analytical modeling and full-wave simulations reveals the evolution of the shear response with twist angle, excitation frequency, layer thickness, and substrate dielectric environment. The interplay between material dissipation and dielectric screening enables continuous regulation of the shear degree and reversible switching between shear-off and shear-on states. Furthermore, we demonstrate a proof-of-concept binary-state encoding scheme based on dissipation-controlled shear-state switching. Our work identifies non-Hermitian dissipation as an additional control parameter in twisted polaritonic systems, expanding the available design space for reconfigurable mid-infrared nanophotonics.

## Introduction

Hyperbolic shear polaritons (HShPs) represent a distinctive class of anisotropic polaritonic states characterized by tilted isofrequency contours (IFCs), asymmetric momentum-space dispersion, and strongly directional subwavelength propagation[1–4]. These features originate from the hybridization of anisotropic polaritonic modes and lead to asymmetric momentum-space dispersion and real-space energy flow. In naturally low-symmetry polar crystals, such shear responses arise from non-orthogonal optical axes and off-diagonal dielectric-tensor components imposed by the crystal lattice[2–6]. Twist optics provides another route, where rotational misalignment between anisotropic layers modifies their relative optical-axis orientation and promotes polaritonic mode hybridization[7–12]. In such twisted structures, geometric and spectral parameters provide additional degrees of freedom for tailoring the shear response[13–15]. Despite their different physical implementations, these systems can be described within a common picture of hybrid polaritonic states formed through non-orthogonal interactions between anisotropic modes. Because realistic polaritonic systems are inherently open and lossy, such hybrid states are generally embedded in a non-Hermitian electromagnetic environment, suggesting that dissipation may provide an additional degree of freedom for tailoring their shear response[16–19].

Material dissipation is commonly associated with resonance broadening, reduced propagation length, and weakened spatial visibility of highly confined polaritonic modes[20,21]. In a non-Hermitian eigenproblem, however, its role is not limited to attenuation. When interacting polaritonic modes experience unequal losses, their relative complex amplitudes and phases can be redistributed, thereby modifying the composition of the hybrid states[22–24]. For anisotropic modes associated with different momentum-space channels, such non-Hermitian reconstruction can alter both the resulting dispersion and the spatial propagation pattern. Dissipation can therefore influence not only how rapidly a polariton decays, but also the degree of asymmetry of the hybrid polaritonic state. This picture motivates treating material dissipation as an additional non-Hermitian control degree of freedom for tuning the shear characteristics of HShPs.

To evaluate the contribution of dissipation to shear control, its influence should be distinguished, as far as possible, from geometric effects that can also modify the hybrid polaritonic response. In twisted anisotropic heterostructures, the relative dispersions of the constituent layers can influence interlayer polaritonic hybridization and reshape the resulting IFCs, thereby contributing to the shear response[8,13,25]. A suitable platform for examining dissipation-controlled shear should therefore minimize such dispersion mismatch while retaining a sufficiently large contrast in material loss. This condition is nontrivial because dispersion and loss both originate from the complex dielectric response of the constituent materials and are generally intertwined. Identifying anisotropic polaritonic materials with closely matched real-part IFCs but substantially different dissipative responses is therefore essential for isolating the influence of modal loss imbalance on hybridized HShPs.

Here, we demonstrate a non-Hermitian route toward HShPs through dissipation engineering in twisted van der Waals heterostructures. Under selected frequency and thickness conditions, the constituent layers exhibit closely matched polaritonic IFCs while retaining a pronounced contrast in dielectric loss, reducing the contribution associated with IFCs mismatch while preserving a substantial dissipative imbalance. An analytical dispersion model, full-wave simulations, and near-field infrared imaging are combined to elucidate the role of dissipation in the formation of HShPs and to examine their dependence on twist angle, excitation frequency, and layer thickness. We further investigate substrate-mediated control of the shear response and the interplay between dielectric screening and material dissipation. Finally, by jointly engineering dissipation and the substrate dielectric environment, we demonstrate controllable switching between shear "on" and shear "off" propagation states and illustrate a proof-of-concept binary-state encoding scheme. These results show that material dissipation can serve as an additional non-Hermitian degree of freedom for tuning the shear characteristics of HShPs in twisted heterostructures.

## Results

### *Unified framework for HShPs mode hybridization*

HShPs originate from the hybridization of anisotropic polaritonic eigenmodes with broken orthogonality. As illustrated in Fig. 1a, two uncoupled eigenmodes (Mode 1 and Mode 2) with distinct eigenfrequencies and gain/loss characteristics can hybridize through an effective coupling term g. Depending on the physical origin, $g$ can be classified into three categories: intrinsic crystal coupling ($g_{\text{crystal}}$), geometric coupling induced by IFCs mismatch ($g_{\text{geometric}}$), and dissipation-mediated non-Hermitian interaction ($g_{\text{non-Hermitian}}$). When the coupled modes possess inequivalent dissipation responses, the hybridization modifies the mode coefficients and spatial field distributions, resulting in asymmetric eigenmode profiles and directional propagation characteristics. In realistic polaritonic systems, the dielectric response is inherently complex, and material dissipation introduces an additional non-Hermitian degree of freedom beyond conventional structural symmetry breaking. To clarify the physical origin of dissipation-induced shear polaritons, we describe the coupled polaritonic modes using an effective non-Hermitian Hamiltonian[18,26,27]:

$$H = \begin{pmatrix} \omega_1 + i\gamma_1 & g \\ g & \omega_2 + i\gamma_2 \end{pmatrix} \tag{1}$$

where $\gamma_1$ and $\gamma_2$ represent the dissipation of the coupled modes. Unlike Hermitian systems with identical loss channels, a dissipation imbalance modifies the hybridized eigenstates and redistributes the modal amplitudes between the coupled polaritonic modes. This non-Hermitian mode reconstruction breaks the symmetry of electromagnetic field distributions and leads to asymmetric momentum-space dispersion. Because the dissipation rate is directly associated with the imaginary part of the dielectric tensor, engineering dielectric loss provides an additional degree of freedom for controlling polaritonic eigenstates beyond conventional approaches based on real-valued dielectric anisotropy.

Although the microscopic origin of the coupling differs among different systems, all previously reported HShPs can be broadly interpreted within a framework of non-orthogonal coupling between anisotropic polaritonic modes. Different physical perturbations, including intrinsic lattice asymmetry, geometric mismatch, and non-Hermitian dissipation, provide distinct coupling channels to break the symmetry of polaritonic eigenstates. Within this unified picture, different physical mechanisms can be classified according to the origin of the mode coupling. Early experimental demonstrations of HShPs were achieved in intrinsically low-symmetry monoclinic polar crystals, where lattice asymmetry naturally introduces non-zero off-diagonal permittivity elements (Fig. 1b). These crystal structural anisotropies provide the coupling terms required for non-orthogonal interaction between phonon polariton modes and generate shear responses[28,29]. However, the limited availability of low-symmetry polar crystals and their intrinsically fixed phonon properties significantly restrict the tunability and reconfigurability of HShPs. To overcome the limitations of intrinsic low-symmetry materials, artificial symmetry engineering based on twisted van der Waals heterostructures has been developed. By introducing controllable rotation and crystal thickness mismatch between anisotropic crystals, the optical axes of adjacent layers are no longer collinear, thus forming mismatched IFCs and breaking mirror symmetry in momentum space (Fig. 1c). This twist-induced mode mismatch enables non-orthogonal coupling between originally orthogonal phonon polariton modes, effectively reproducing the shear response of low-symmetry crystals and allowing customizable HShPs[15,30].

However, these above approaches rely on structural symmetry breaking, where geometric or crystallographic parameters determine the coupling between polaritonic modes. Here, we introduce a distinct non-Hermitian pathway for generating shear polaritons through dissipation engineering (Fig. 1d). In contrast to previous mechanisms where the effective coupling originates from crystal anisotropy or twist-induced IFCs mismatch, here the effective interaction responsible for asymmetric hybridization is mediated by the non-Hermitian response of the system. Non-Hermitian optical responses do not require balanced gain and loss and can also be realized in purely passive, loss-engineered systems[18,24]. In rationally designed twisted heterostructures, shear polaritons can emerge even when the IFCs of the constituent layers are identical. This phenomenon originates from material dissipation acting as an active non-Hermitian degree of freedom. The imaginary part of the dielectric response produces direction- and layer-dependent attenuation, which modifies the relative complex amplitudes of the coupled polaritonic components[18,26,31]. The real part of the dielectric response predominantly determines the dispersion, whereas its imaginary part controls attenuation and can modify the complex composition of hybrid modes. Consequently, even with matched constituent-layer IFCs at a fixed twist geometry, unequal dielectric losses between constituent layers can induce asymmetric complex eigenmodes and break the orthogonality of polaritonic fields. By engineering material dissipation, the degree of non-Hermitian modal asymmetry can be manipulated to generate pronounced shear effects without modifying geometric parameters. This dissipation-driven mechanism establishes a new route toward actively controlled shear polaritonics.

To elucidate the role of dissipation in shaping the dispersion and shear characteristics of HShPs in twisted heterostructures, we employ Maxwell's equations to analyze the polaritonic dispersion of the system. We consider a configuration consisting of two in-plane anisotropic material layers stacked on a substrate, modeled as semi-infinite in the lateral directions. In our analysis, the finite thicknesses of the individual layers are taken into account, with the top layer thickness denoted as $d_1$ and the bottom layer thickness as $d_2$. Both the air superstrate and the substrate are treated as infinitely thick. The analytical expression for the IFCs of HShPs takes the form of the following nonlinear dispersion equation (for detailed derivation, see Supplementary

Note 1):

$$k\rho_1(\varphi-\theta)d_1+\phi_{\text{air}}+\phi_{12}+m\pi=0 \quad (2)$$

Here, $\varphi$ denotes the angle between the wavevector $k$ and the x-axis, $\theta$ represents the angle between the [100] crystallographic direction of the top-layer material and the x-axis and m is an integer. $\phi_{\text{air}}$ and $\phi_{12}$ are the reflection phases at the material 1–air interface and the material 1–material 2 interface, respectively. Owing to the complexity of the full expression, the complete form of the dispersion equation is provided in Equations 13 and 14 of Supplementary Note 1. As shown in Fig. 1e, the IFCs of HShPs supported in the twisted bilayer α-$MoO_3$ structure were calculated using Equation (2). For conventional hyperbolic polaritons, the IFCs exhibit a symmetric hyperbolic shape with an optic axis angle of $\alpha = 0$ and equal polar angles $\varphi_1 = \varphi_2$. Consequently, both the real part (Re($k$)) and imaginary part (Im($k$)) of the wavevector also display symmetric characteristics (Supplementary Fig. 3). In contrast, HShPs possess an inherently asymmetric nature. The results in Fig. 1e demonstrate that the IFCs of the twisted structure exhibit a non-zero optic axis angle ($\alpha \neq 0$) and unequal polar angles ($\varphi_1 \neq \varphi_2$).

The shear characteristics of HShPs can be tuned not only by the thickness and twist angle of the bilayer materials but also further regulated through the dissipation of the system. For simplicity, this work focuses on modulating Im($\varepsilon_x$) of material 2 to demonstrate that system dissipation can effectively control HShPs. This tunability originates from the system dissipation's ability to govern the asymmetric energy dissipation of HShPs. Here, a parameter denoted as $\kappa = e^{-2\pi[\text{Im}k/\text{Re}k]}$ is introduced to characterize this asymmetric energy dissipation behavior. This parameter quantifies the extent to which the amplitude of HShPs decays to $\kappa$ times its original value over one wavelength. We calculated the values of κ for different Im($\varepsilon_x$) (Im($\varepsilon_x$) = 0, 0.5, and 2) in Fig. 1f, showing the regulatory effect of system dissipation on the asymmetric energy dissipation of HShPs. As Im($\varepsilon_x$) increases, a pronounced enhancement in the decay of $\kappa$ is clearly observed. To intuitively illustrate the regulation of HShPs in the heterostructure by system dissipation, we calculated the IFCs of HShPs for different values of Im($\varepsilon_x$) in Fig. 1g, demonstrating that Im($\varepsilon_x$) can be utilized to control the shear polaritons induced by twisting in the heterostructure. Increasing Im($\varepsilon_x$) can effectively enhance the non-Hermiticity and eigenmode asymmetry of the system, thereby providing an extra degree of freedom for manipulating HShPs in twisted structures.

***Realization of HShPs in twisted bilayer heterostructures via dissipation engineering***

To verify the existence of HShPs arising from the proposed mechanism, we designed a twisted heterostructure composed of α-$MoO_3$ and α-$V_2O_5$ (Fig. 2a). Both materials are orthorhombic anisotropic crystals supporting phonon polaritons and exhibit distinct dissipative responses. This dissipation mismatch introduces an effective dielectric asymmetry, which enables the emergence of a pronounced shear effect. It can be observed that the IFCs profiles of the two materials exhibit favorable consistency under specific conditions, effectively eliminating the HShPs induced by inconsistent IFCs (Fig. 2b). To quantitatively characterize the asymmetry of HShPs in the heterostructure as a function of excitation frequency and twist angle, a dimensionless parameter defined as the asymmetric factor (AF = $\eta_1/\eta_2$) is introduced (see Supplementary Note 2 for details). Under this definition, when the AF = 1, the polariton distribution in momentum k space remains fully symmetric. In this case, no redistribution of dissipation occurs within the IFCs, and the dispersion retains the conventional hyperbolic profile (Supplementary Fig. 2a). For clarity in the following discussion, we take AF = 1 as the reference value; deviations from this baseline directly reflect the strength of the shear effect, with larger deviations indicating more pronounced asymmetry in the polaritonic response.

The dependence of the AF on excitation frequency and twist angle is summarized in Fig. 2c and Fig. 2d. At fixed twist angles, the spectral evolution of the AF shows distinct trends (Fig. 2c). At 20°, the AF increases progressively with excitation frequency, whereas at 30° and 45° it first decreases near 890 $cm^{-1}$ and subsequently increases toward higher frequencies. At fixed excitation frequencies, the angular dependence is likewise non-monotonic (Fig. 2d). At 700 $cm^{-1}$, the AF decreases with increasing twist angle up to 30° and then recovers toward unity at 45°. At 880 and 900 $cm^{-1}$, the AF initially increases with twist angle and subsequently decreases, with the maximum appearing near 30° and 20°, respectively. These trends indicate that the shear response depends sensitively on both excitation frequency and twist configuration.

These results further confirm the existence of loss-induced shear effects in twisted configurations. Moreover, this analysis highlights a key advantage of twisted heterostructures: the ability to precisely modulate and enhance shear polaritons simply by adjusting the twist angle and excitation frequency. In contrast, the shear response of conventional bulk monoclinic crystals is rigidly fixed by the intrinsic crystal structure, precluding such flexible and efficient control. We also calculated the IFCs of HShPs supported in the twisted bilayer α-$MoO_3$ structure and those in the α-$MoO_3$/α-$V_2O_5$ structure. The shear branches in the heterostructure exhibit a faster energy decay than those in twisted bilayer α-$MoO_3$ (Supplementary Fig. 4).

To experimentally characterize the artificially engineered shear polaritons in twisted α-$MoO_3$/α-$V_2O_5$ heterostructures, we fabricated two samples with different twist angles (30° and 45°), as shown in Fig. 2e and 2f. The blue dots in the figures indicate point defects that serve as excitation sources for the near-field measurements. Near-field imaging measurements were then performed using scattering-type scanning near-field optical microscopy (s-SNOM) on these heterostructures. At a twist angle of 30° and an excitation frequency of 900 $cm^{-1}$, a pronounced shear effect emerges in the heterostructure. The experimentally observed near-field amplitude image (Fig. 2g) exhibits the characteristic asymmetric wavefront of HShPs, which is further confirmed by the corresponding Fast Fourier transform (FFT) spectra (Fig. 2k). The FFT of the near-field distribution provides direct access to the momentum-space polaritonic dispersion and is widely used to reconstruct IFCs in polaritonic systems[32–34]. The observed wavefronts and Fourier-space features are consistent with the corresponding simulations (Fig. 2h and 2l). When the twist angle is increased from 30° to 45°, similarly pronounced asymmetric shear polaritonic wavefronts are observed at the same excitation frequency (900 $cm^{-1}$). The experimental results (Fig. 2i and Fig. 2m) are in good agreement with the corresponding simulations (Fig. 2j and Fig. 2n), further validating the existence of shear polaritons. The propagation characteristics of HShPs are further tunable through external excitation conditions. Systematic near-field measurements reveal that the asymmetric shear polaritonic wavefronts remain robust over a broad range of excitation frequencies and polarization configurations, confirming the intrinsic stability of dissipation-driven shear states in twisted heterostructures. The detailed frequency-dependent and polarization-dependent evolutions of HShPs are provided in Supplementary Notes 3 and 4.

Although substrate dielectric environments are known to influence polariton dispersion[35–38], their role in regulating the shear response of HShPs in twisted heterostructures remains unclear. We therefore investigate the substrate dependence of HShPs in twisted α-$MoO_3$/α-$V_2O_5$ heterostructures (Fig. 3a). The dispersion relation was initially obtained through Equation (2) by incorporating different substrate dielectric constants ($\varepsilon_{sub}$ = -2, -1, 0, 2, 5). This yielded the real part Re(k) and imaginary part Im($k$) of the wavevector for HShPs in the twisted heterostructure under various substrate dielectric environments, as shown in Fig. 3b and Fig. 3c. We found that variations in $\varepsilon_{sub}$ significantly affect the distributions of both Re($k$) and Im($k$). However, they do not eliminate the shear effect: in all cases, Re($k$) exhibits an asymmetric distribution, with a peak appearing in the negative polar angle range but absent in the positive polar angle range. For Im($k$), which typically represents the propagation loss of HShPs (where Im($k$) = 0 corresponds to a lossless propagation scenario), the curve for $\varepsilon_{sub}$ = 0 (blue curve in Fig. 3c) is markedly lower than those for other $\varepsilon_{sub}$ values. This indicates that the substrate dielectric environment can effectively regulate the propagation loss of HShPs and enable HShPs with low propagation loss. Leveraging the low propagation loss of HShPs at $\varepsilon_{sub}$ = 0, we performed near-field simulations for this case (Fig. 3d). The resulting HShPs displays a ray-like propagation feature, which is distinctly different from the fan-shaped propagation characteristic typically observed on conventional $SiO_2$ substrates. The corresponding FFT image (Fig. 3h) also exhibits features distinct from those on $SiO_2$ substrates, with the upper and lower branches of the minima both appearing in the low-momentum region (approximately $1.2k_0$).

Based on the predicted response near $\varepsilon_{sub} = 0$, we selected sapphire ($Al_2O_3$), which exhibits a near-zero dielectric response around 900 $cm^{-1}$, as the supporting substrate[39]. We fabricated twisted heterostructure samples on a sapphire substrate with twist angles of 30°, 45°, and 60°, as shown in Supplementary Fig. 9. High-resolution near-field imaging was performed using s-SNOM. The experimental results reveal that the asymmetric polaritonic response is preserved across the investigated twist angles and substrate conditions. At 900 $cm^{-1}$, the real-space images clearly display the characteristic asymmetric propagation wavefronts of shear polaritons. Both the near-field images (Fig. 3e–g) and their corresponding FFT spectra (Fig. 3i–k) show pronounced asymmetric wavefronts and momentum-space distributions, indicating that sapphire, as a low-loss dielectric substrate, preserves the shear response within the heterostructure. Notably, compared with $SiO_2$ substrates, sapphire induces a distinct modulation of the polariton dispersion. This difference arises from its unique dielectric function and phonon resonance characteristics, which significantly alter the electromagnetic boundary conditions of the system. As a result, polariton propagation that exhibits conventional hyperbolic wavefronts on $SiO_2$ evolves into highly localized and discrete ray-like propagation patterns on sapphire. The emergence of these highly collimated features highlights the critical role of substrate-induced momentum-space reshaping of hyperbolic modes and provides direct experimental evidence for substrate engineering as a powerful approach to control nanoscale light propagation and information routing.

To further investigate the role of the substrate dielectric environment in regulating the non-Hermitian shear response, systematic full-wave simulations were performed with different substrates (Supplementary Fig. 10). At 900 $cm^{-1}$, HShPs exhibit strong environmental robustness on typical dielectric substrates, including SiC, $LiV_2O_5$, AlN, and $SrTiO_3$, where asymmetric wavefronts are well preserved despite variations in dielectric constants. This indicates that moderate dielectric perturbations have limited influence on the interlayer hybridization responsible for shear polariton. In contrast, replacing the substrate with a metallic Au layer leads to the suppression of asymmetric wavefronts and the disappearance of the shear response. This behavior originates from the strong electromagnetic screening and modified boundary conditions introduced by the metal substrate, which drive the system toward highly confined high-momentum modes with reduced interlayer field overlap[4,40,41]. As a result, the hybridization between twisted layers is weakened, suppressing the non-Hermitian coupling required for HShPs. These results reveal that the shear response is governed by the competition between interlayer non-Hermitian coupling and electromagnetic confinement, providing an additional route for externally regulating HShPs states.

***Non-Hermitian dissipation engineering enables reversible switching of HShPs states***

Previous approaches for manipulating HShPs mainly rely on geometric parameters, excitation conditions, or external dielectric environments, which can modify polaritonic distributions but offer limited capability for reversible switching between symmetric and asymmetric propagation states. Here, we combine dissipation and substrate engineering to achieve reversible switching of the shear state without altering the twist geometry. Fig. 4a illustrates the corresponding s-SNOM configuration, in which the polaritonic response can be switched between a nearly symmetric ("off") state and an asymmetric ("on") state through joint control of material dissipation and substrate dielectric screening.

The role of dissipation engineering in modulating HShPs within the twisted heterostructure is first examined (Fig. 4b). Specifically, near-field distributions are calculated for a series of $\mathrm{Im}(\varepsilon_x)$ values, and the corresponding HShPs amplitude profiles are extracted along a prescribed path, enabling a quantitative characterization of near-field asymmetry. As $\mathrm{Im}(\varepsilon_x)$ increases, both peak 1 ($|E_z^{\mathrm{I}}|$) and peak 2 ($|E_z^{\mathrm{II}}|$) exhibit a monotonic decay, which can be attributed to enhanced dissipative loss in the system. The peak-amplitude difference, $\Delta|E_z|$, varies non-monotonically with $\mathrm{Im}(\varepsilon_x)$, increasing in the low-dissipation regime and decreasing at higher loss levels, with a maximum at $\mathrm{Im}(\varepsilon_x) = 1.2$. This behavior indicates the existence of an optimal dissipation window for maximizing shear-induced asymmetry in HShPs. Excessive loss shortens the polariton propagation length and reduces the spatial visibility of the hybrid wavefront, consistent with the general limitations of observing non-Hermitian modal effects in strongly dissipative passive system[42]. Therefore, an intermediate dissipation regime provides optimal conditions for maximizing asymmetric mode hybridization. We next examine the combined influence of $\varepsilon_{sub}$ and material dissipation on the shear response. Numerical simulations show that the Im($\varepsilon_x$) associated with a given shear criterion shifts to higher values as the $\varepsilon_{sub}$ increases, exhibiting an approximately linear dependence (Fig. 4c). This trend suggests a competitive relationship between dielectric screening and dissipation in determining the shear response. At a fixed $\varepsilon_{sub}$, increasing dissipation within an appropriate range enhances the field asymmetry, whereas at a fixed dissipation level, increasing $\varepsilon_{sub}$ progressively reduces the asymmetric field distribution (Supplementary Fig. 11).

To evaluate its performance, comprehensive numerical simulations were conducted, and the AF was extracted systematically. For visualization, we classify states with $0.95 \leqslant \mathrm{AF} \leqslant 1.05$ as nearly symmetric ("shear-off") and states outside this interval as asymmetric ("shear-on"). As illustrated in Fig. 4d, the shear state of the hyperbolic polaritons within the heterostructure is clearly governed by the dissipation status, enabling robust on/off switching. Capitalizing on this controllable switching behavior, we illustrate a proof-of-concept binary-state encoding scheme (Fig. 4e). By modulating the dissipation status, the system toggles between the on and off shear states, demonstrating the potential of shear-polariton states for binary optical encoding and reconfigurable nanophotonic information processing[43].

**Conclusions**

In summary, we establish dissipation-mediated non-Hermitian interaction as a distinct route for realizing and controlling hyperbolic shear polaritons in twisted α-$MoO_3$/α-$V_2O_5$ heterostructures. Twist-induced hybridization provides the underlying platform for shear polariton formation, while dielectric-loss contrast introduces an additional non-Hermitian degree of freedom for regulating the complex hybrid polaritonic states and their momentum-space asymmetry. Near-field infrared imaging, analytical modeling, and full-wave simulations consistently reveal the evolution of the shear response with twist angle, excitation frequency, layer thickness, and substrate dielectric environment. Beyond geometric tuning, the interplay between material dissipation and dielectric screening enables continuous modulation of the shear degree and reversible switching between shear-off and shear-on states. These switchable states further support a proof-of-concept binary-state encoding scheme. By integrating twist optics with dissipation control, this work establishes twisted anisotropic heterostructures as a versatile platform for tailoring shear polaritons and provides an additional route toward reconfigurable mid-infrared polaritonic functionalities.

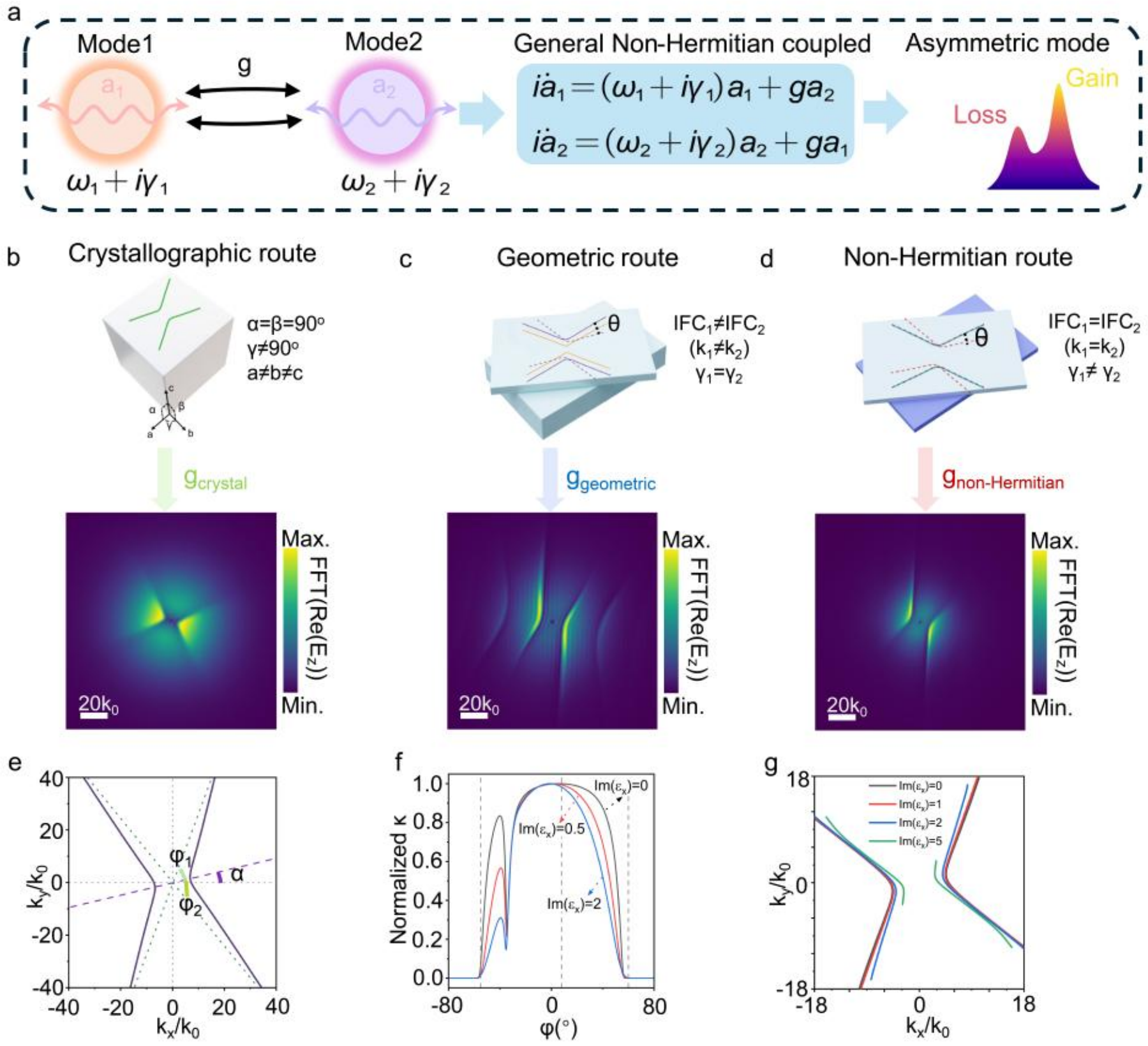

**Fig. 1. Unified framework for symmetry breaking and dissipation engineering of hyperbolic shear polaritons. a.** Schematic diagram of the principle of asymmetric modes generated by non-Hermitian coupled modes. **b.** HShPs induced by crystal anisotropy mediated coupling. Taking the monoclinic crystal system as an example, the upper panel shows a schematic of the bulk monoclinic crystal, and the lower panel presents the Fast Fourier-transformed (FFT) Re($E_z$) image of hyperbolic shear polaritons in the monoclinic crystal. **c.** HShPs induced by IFCs mismatch mediated geometric coupling. Taking twisted bilayer α-$MoO_3$ with different thickness as an example, the upper panel shows a schematic of the twisted bilayer structure, the two layers exhibit differently shaped IFCs, and the lower panel displays the FFT Re($E_z$) image of hyperbolic shear polaritons in the twisted bilayer. **d.** HShPs induced by dissipation mediated non-Hermitian interaction. Taking twisted bilayer α-$MoO_3$ with identical thickness as an example, the constituent layers are assigned matched IFCs but different dielectric losses. The upper panel shows a schematic of the twisted bilayer structure, and the lower panel displays the FFT Re($E_z$) image of hyperbolic shear polaritons in the twisted bilayer. **e.** IFCs of hyperbolic shear polaritons exhibiting a non-zero optic axis angle ($\alpha \neq 0$) and unequal polar angles ($\varphi_1 \neq \varphi_2$). **f.** Analytically calculated κ for different values of Im($\varepsilon_x$). **g.** Analytically calculated IFCs for different values of Im($\varepsilon_x$).

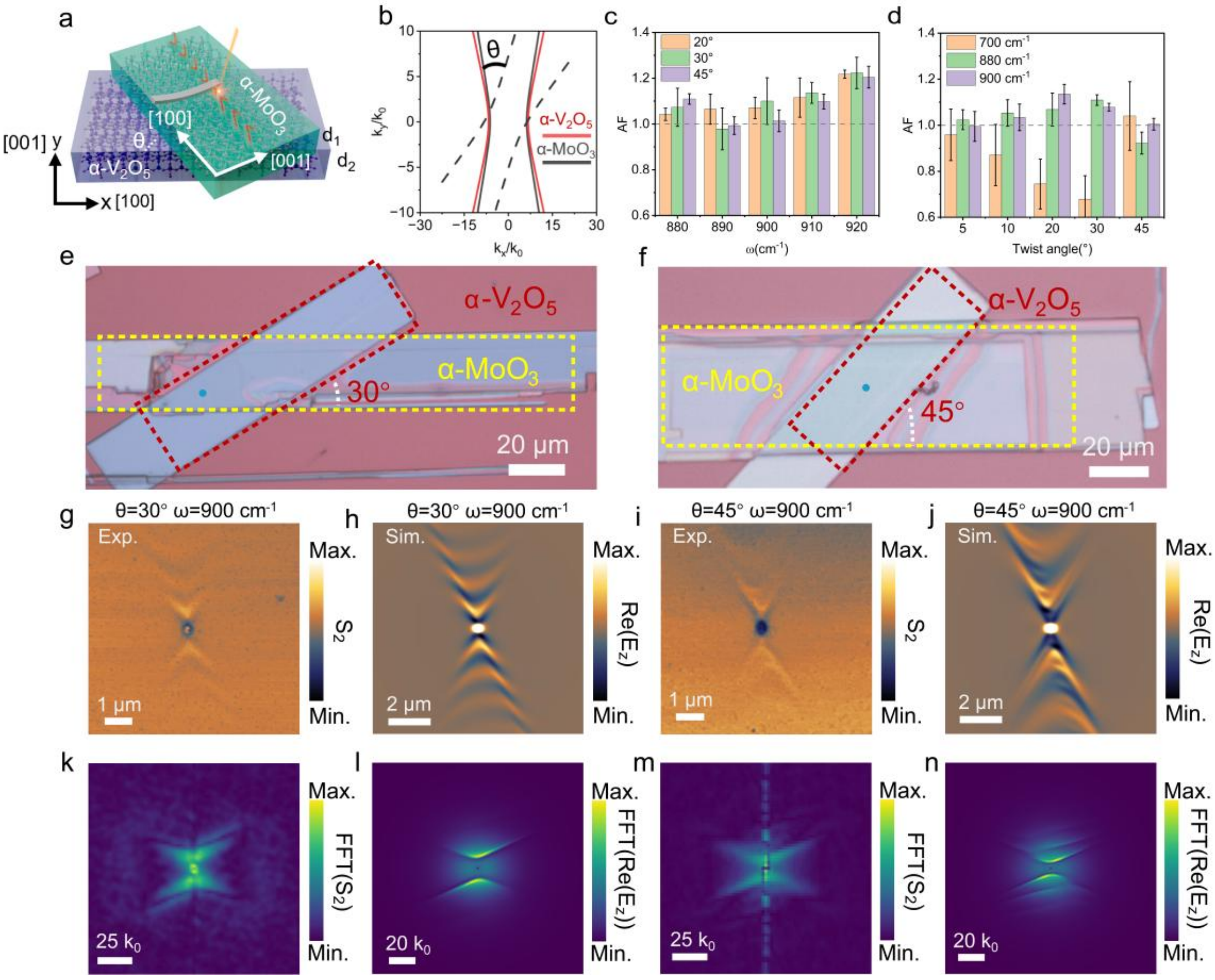

**Fig. 2. Experimental observation of dissipation-driven HShPs in twisted heterostructures. a.** Schematic of infrared scattering-type scanning near-field optical microscopy (s-SNOM) measurements performed on a twisted bilayer heterostructure. The structure consists of two layers: the top layer is $\alpha$-$MoO_3$ with thickness $d_1$ and the bottom layer is $\alpha$-$V_2O_5$ with thickness $d_2$. The x- and y-axes align with the [100] and [001] crystallographic directions of the bottom layer, respectively. A twist angle $\theta$ exists between the two layers. **b.** IFCs of $\alpha$-$MoO_3$ and $\alpha$-$V_2O_5$ at an excitation frequency of 900 $cm^{-1}$. The thickness of $\alpha$-$MoO_3$ is 150 nm and that of $\alpha$-$V_2O_5$ is 200 nm. **c.** Bar plots of the numerically simulated HShPs asymmetric factor (AF) as a function of excitation frequency at twist angles of 20°, 30°, and 45°, respectively. In the model, the top layer thickness is 150 nm and the bottom layer thickness is 200 nm. **d.** Bar plots of the numerically simulated HShPs AF as a function of twist angle at excitation frequencies of 700, 880, and 900 $cm^{-1}$, respectively. In the model, the top layer thickness is 150 nm and the bottom layer thickness is 200 nm. **e, f.** Optical microscope images of samples on $SiO_2$ at different twist angles (30° and 45°) **g, i.** Experimental near-field images of HShPs at an excitation frequency of 900 $cm^{-1}$ and twist angles of 30° and 45°, respectively. **k, m.** Experimental FFT images corresponding to panels **g** and **i**, respectively. **h, j.** Simulated near-field images of HShPs at an excitation frequency of 900 $cm^{-1}$ and twist angles of 30° and 45°, respectively. **l, n.** Simulated FFT images corresponding to panels **h** and **j**, respectively.

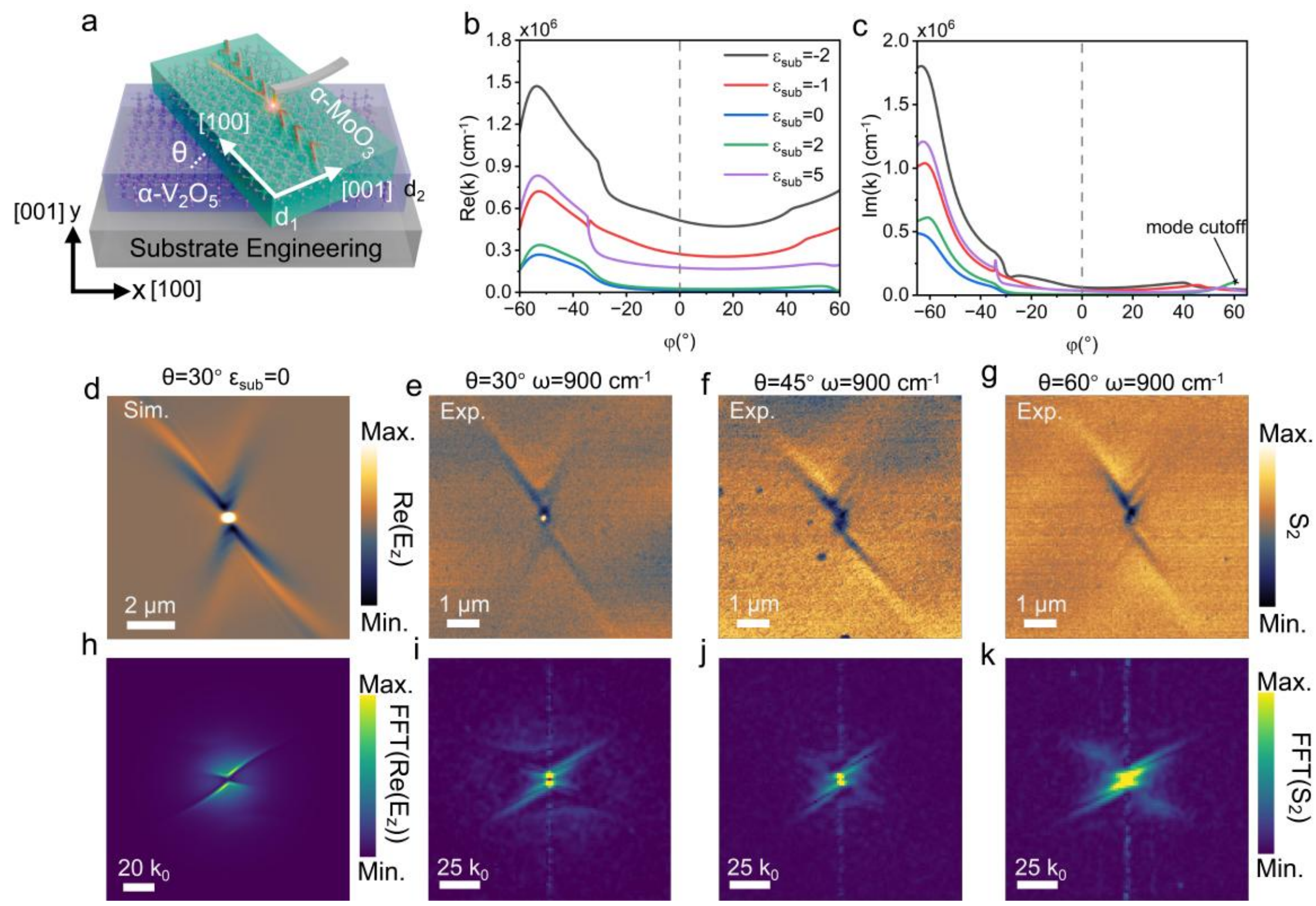

**Fig. 3. Substrate engineering for tuning dissipation-driven HShPs in twisted heterostructures. a.** Schematic of s-SNOM measurements for substrate-engineered tuning of HShPs in a twisted heterostructure. The structure consists of three layers: the top layer is α-$MoO_3$, the middle layer is α-$V_2O_5$, and the bottom layer is the substrate engineering material. The x- and y-axes correspond to the [100] and [001] crystallographic directions of the bottom layer, respectively. The thicknesses of the α-$MoO_3$ and α-$V_2O_5$ layers are $d_1$ and $d_2$, respectively, with a twist angle θ between the two van der Waals layers. **b, c.** Analytically calculated real part (Re(k)) and imaginary part (Im(k)) of the wavevector as a function of polar angle φ for different substrate permittivity ($\varepsilon_{sub}$), respectively. **d.** Simulated near-field image of HShPs in the heterostructure at an excitation frequency of 900 cm$^{-1}$, twist angle of 30°, and $\varepsilon_{sub} = 0$. **h.** Simulated FFT image corresponding to panel **d**. **e–g.** Experimental near-field amplitude images of HShPs wavefronts measured by s-SNOM at an excitation frequency of 900 cm$^{-1}$ and twist angles of 30°, 45°, and 60°, respectively. The substrate is sapphire. $S_2$ denotes the second-harmonic near-field signal. **i–k.** Experimental FFT images corresponding to panels **e–g**, respectively.

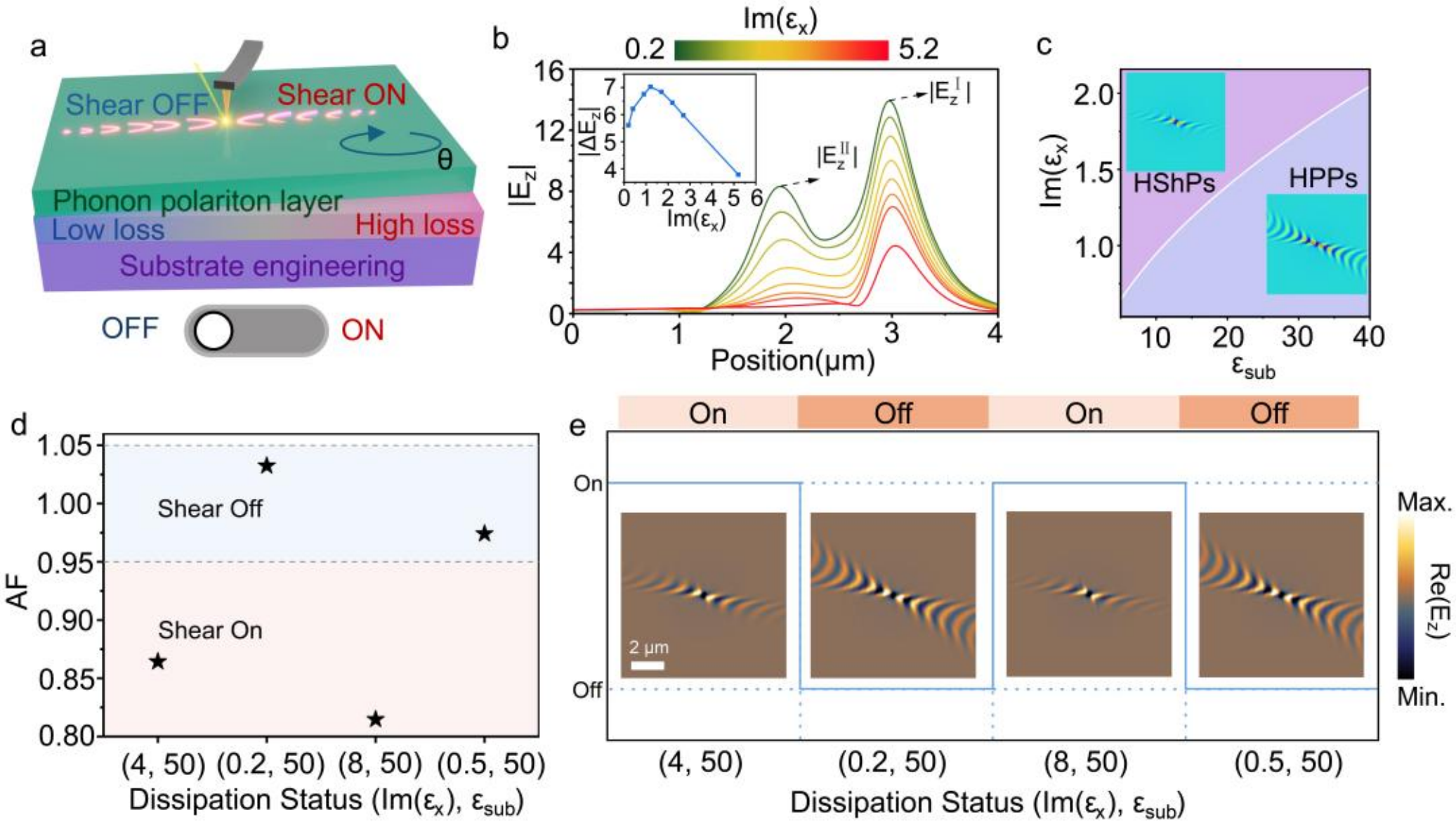

**Fig. 4. On/off switching of HShPs in twisted heterostructures via dissipation and substrate engineering. a.** Schematic of s-SNOM measurements for HShPs in a twisted heterostructure under combined substrate and dissipation engineering. The structure is divided into three layers. The first layer is the phonon polariton layer; the second layer is the dissipation engineering layer, with the left side being the low-loss region and the right side being the high-loss region; and the third layer is the substrate engineering layer. **b.** Imaginary part of the x-direction permittivity ($\mathrm{Im}(\varepsilon_x)$) of different dissipation-engineered materials and the corresponding extracted $E_z$ amplitude profiles along a specified line, showing peak 1, peak 2, and their difference. The main panel shows the extracted $|E_z|$ amplitude distributions for $\mathrm{Im}(\varepsilon_x)$ = 0.2, 0.4, 0.9, 1.2, 1.7, 2.2, 2.7, and 5.2. The inset displays the difference between Peak 1 and Peak 2 in the $|E_z|$ amplitude distributions. **c.** Critical $\mathrm{Im}(\varepsilon_x)$ values required for the observation of pronounced HShPs as a function of $\varepsilon_{sub}$. The purple region indicates the HShPs regime, the light purple region indicates the hyperbolic phonon polaritons (HPPs) regime, and the white curves represent the fitting curves of the simulation results. **d.** On/off control of the shear state via dissipation modulation and substrate engineering. The "shear off" state is represented by AF values falling within the 0.95–1.05 interval, while all other ranges signify the "shear on" state. **e.** On/off states of HShPs under various dissipation status along with their corresponding simulated near-field distribution profiles.